\def\vec#1{{\rm\bf #1}}
\def\lz{\ell_{\parallel}}
\def\lp{\ell_{\bot}}
\def\matr#1{\bbox{\breve{#1 \,}}}

\documentstyle[prabib,aps]{revtex}

\begin{document}
\draft
\title{Non-uniform deformations in liquid crystalline elastomers.}

\author{ E.M. Terentjev, M. Warner and G.C. Verwey}

\address{Cavendish Laboratory, Madingley Road, Cambridge,
CB3 0HE, U.K. }

\maketitle

\begin{center}
{\bf Abstract}
\end{center}

\begin{abstract}
We develop a molecular model of non-uniform deformations
within the framework of liquid crystal rubber elasticity.
We show that, similarly to the uniform case, the theory is
not sensitive to the molecular details of polymer liquid crystal
involved and the resulting elastic free energy is quite
universal.  The result of this work is the general
expression  for the free energy of deformations, which combines
the effects of large non-symmetric affine strains in the rubbery
network and gradients of curvature deformations of the director
field, $F  \sim \lambda^T \{ \nabla {\bf n} \}^2 \lambda $. We
derive the molecular expressions for the 5 principal
independent elastic constants governing non uniform deformations in
the presence of elastic strains. They
also depend on the polymer step length anisotropy and - most
strikingly - have an overall \underline{negative} sense.  We
therefore predict that in some circumstances, especially at large
elastic deformations $\lambda$, these new terms may overpower the
usual, positive Frank elastic moduli of the underlying nematic
structure, as well as the coupling in nematic elastomers of
uniform relative rotations of the director and the elastic
matrix.  In this event highly distorted polydomain textures {\bf
n}({\bf r}) would be favoured.

\end{abstract}

\vspace{0.2in}

\pacs{PACS: 61.40K - 61.30B - 46.30C}


\widetext

\section{Introduction}
A large amount of experimental and theoretical work
has been invested in recent years into studies of liquid
crystalline elastomers and gels. Elastomers (as opposed to
hard and brittle resins) are weakly crosslinked,
percolating networks of polymers chains which retain a
significant molecular mobility of their strands. Rubber,
therefore, is a material with a very low shear modulus
$\mu$ (about $10^{-4} - 10^{-5}$ times that of conventional
elastic solids) and thus deforms at practically constant
volume. Its weakness is matched by its capacity to
withstand very large deformations, commonly 100's \% and
more, thus allowing large amounts of elastic energy to be
stored in the material.

Nematic liquid crystals (and liquid crystal polymers) also
represent an intermediate state between conventional liquids and
solids. They flow under applied stress, but they have a long range
orientational order and thereby a curvature elasticity associated
with deformations of this order. When these two remarkable
physical systems are combined in the same material, {\it i.e.} a
weakly crosslinked network is formed of mesogenic polymer chains
capable of spontaneous orientational ordering, a
qualitatively new state of matter emerges. Elastomers, being
nominally solids, display a high molecular mobility since
connected crosslinks are distant from each other and the chains of
mesogenic monomers have much freedom. The axis of
orientational symmetry breaking (the nematic director \vec{n}) is
mobile too and responds to imposed elastic strains. We have,
therefore, a uniaxial solid where the preferred direction can be
altered when the body is elastically deformed; \vec{n} becomes
an independent elastic variable. Such systems resemble the
so-called Cosserat medium, where the
internal torques are allowed and, therefore, elastic stress can
be non-symmetric.

For most deformations \vec{n} is anchored in the
elastic medium by the crosslinking. Its rotation requires the
input of some external work, but at large strains it may jump in
direction. For some deformations the director is (theoretically)
freely mobile, although still intimately coupled to the network,
and deformations occur without resistance, that is without
applied stress.  Thus nematic elastomers seem to exhibit a
qualitatively different elasticity from other solids and, in the
non-linear regime, display nematic and mechanical transitions
hitherto unknown. Elastic non-linearity is an important motivation
for developing a molecular theory of nematic elastomers: rubbery
networks are capable of extremely large extensions and continuum
elastic models are bound to break down well before elastomers do.

Another important motivation is that a specific picture of
{\it how} the mesogenic chains spontaneously change their
equilibrium shape on entering the nematic phase is
unnecessary for a detailed description of most liquid crystal
elastomer properties. This
allows a universal description of nematic elastomers, a
modest extension of the classical molecular theory of
conventional rubbers, independent of molecular models of
liquid crystal polymers where there is much less agreement.

Experimental studies of liquid crystal elastomers  began more
than a decade ago \cite{fink81} and various side-chain
\cite{zent-old,zent-L,mitch-0,fink-L}  and main-chain
\cite{zent-MC,ober12,sirigu} materials have been investigated.
However, it was quickly discovered that if the mesogenic polymer
is crosslinked in the isotropic phase, the nematic phase
obtaining on cooling down through the clearing point $T_{ni}$
inevitably has a scattering, highly nonuniform equilibrium
texture of the director. Although the question of quenched random
disorder  created by misoriented crosslinks  has some fundamental
interest, for most practical effects and applications a uniform
monodomain birefringent elastomer was preferrable. Several
methods of obtaining such systems have been developed, notably
crosslinking in a magnetic field \cite{mitch-mono} and two-step
crosslinking with  stress applied in the intermediate state
\cite{fink-mono}. Experiments then showed the effects of coupling
of the mobile anisotropy axis to the elastic strain field in such
monodomain nematic elastomers, for example the switching of the
director by the perpendicular extension has been observed
\cite{mitch-prl}. Interested reader can find more information in
the recent review articles \cite{ober-review,our-review}.

There has been a substantial effort in developing both the
continuum (applicable only for small deformations) and the full
molecular theory of nematic elastomers, which are also reviewed
in \cite{our-review}. Much of this effort has been
successful in  describing the existing and predicting new
physical effects. However, all these theoretical models
addressed only the problem of uniform deformations and there has
been no adequate theory describing the effect of director
curvature.  At the same time, there are many mechanisms of
director deformation, for example due to surface anchoring,
disclinations, or in transient regimes. Therefore, it is
important to know how director curvature is affected by elastic
deformations of the rubber.  The present paper is devoted to this
particular point: after a brief overview of basic concepts and
results of the classical theory in the next Section, we derive
the full non-uniform nematic rubber elastic free energy, which
depends on the products of elastic strains tensors, chain step
length tensors and on the second power of $\nabla \vec{n}$ (only
in chiral - cholesteric - systems can we expect
 linear gradients of \vec{n} \cite{terent}). In the last Section
we examine the general free energy form via simple examples and
compare its effect with that of standard Frank elasticity. The
main conclusion of this paper is that some of the new non-uniform
rubber elastic terms are \underline{negative} and, in some
circumstances, may favour equilibrium non-uniform textures
\vec{n}(\vec{r}).

\section{Basic theory of uniform nematic elastomers}

The first linear continuum
picture of the local anchoring of the director with respect
to the elastic matrix is due to de Gennes \cite{degen80},
the energy density being (phenomenologically)
\begin{equation}
E_{rot} = \frac{1}{2}b_1 \, [(\bbox{\Omega} -
\bbox{\omega}) \times \vec{n}]^2 + b_2 \ \vec{n} \cdot
\matr{e} \cdot [(\bbox{\Omega} - \bbox{\omega}) \times
\vec{n}] \  . \label{rotations}
\end{equation}
The local rotation of
the elastic medium is $\bbox{\Omega}$, given by the
antisymmetric part of the infinitesimal deformation tensor,
$e^A_{ij}$: \ $\Omega_i = \epsilon_{ijk} e_{jk}$.
$\bbox{\omega}$ is the rotation of the director \vec{n}
about an axis parallel to $\bbox{\omega}$, the change in
the director being for small rotations $\delta \vec{n} =
[\bbox{\omega}\times \vec{n}]$.
The first term in Eq.(\ref{rotations}) expresses the penalty
for the relative director-matrix rotations.  It is quite unlike
usual nematic Frank elasticity, which depends on gradients of
director rotation.  In nematic solids uniform rotations are
also penalised and the degeneracy of the local direction of
orientational order is removed.  The $b_2$ term is the first hint
that it is not only rotations of the anchoring matrix that can
rotate the director, but that the symmetric component of shear
strain is coupled to the director as well.

More recently, molecular models of nematic elastomers
 have been constructed \cite{abram,WGV}, which led to an
understanding of mechanical critical points, memory of
crosslinking,  shifts in phase equilibria  and stress-strain
relations. The essential anisotropy of polymer chains
leads to a straightforward modification of the conventional
Gaussian rubber elasticity theory. The theory can be further
developed, as in conventional networks, to account for
junction point  fluctuations \cite{WGV} and other effects,
but to understand the startling new effects visible in
nematic elastomers these elaborations are not necessary.
The theory is based on a single parameter, the anisotropy of
the polymer strand shape, {\it i.e.} on the ratio of the mesogenic
chain persistence lengths along and perpendicular to the nematic
director, $\lz/\lp$. One should note that this ratio strongly
depends on the molecular nature of the polymer ($\lz/\lp \sim
1.5$ in side-chain liquid crystal polymers
\cite{mitch-neutron,sixou}, while in a main-chain system one can
find $\lz/\lp \sim 10$ or more \cite{blumst}) and that different
values are predicted by the different theoretical approaches
(which include freely-jointed chains, worm-like persistent chains,
rotationally-isomeric chains, etc.).  The happy situation in the
theory of liquid crystal elastomers is that this single parameter
$\lz/\lp$ is directly related to the macroscopic sample shape and
can be easily measured simply by changing temperature through the
nematic-isotropic transition \cite{fink-L} and observing the
spontaneous sample shape change. After such a measurement the
theory does not have any free parameters and should make accurate
predictions.

The basic theory assumes that in a nematic monodomain with
director ${\bf n}^0$, a given polymer chain span between
connected crosslinking points, at the moment of crosslinking,is
${\bf R}^0$  . If this chain
is long enough (so that we obtain an elastomer, not a resin),
this  end-to-end distance has a  Gaussian distribution:
\begin{equation}
P_0({\bf R}^0) \sim
Det[l^0_{ij}]^{-1/2} \exp \left( - \frac{3}{2{\cal L}}R^0_i
(\ell^0_{ij})^{-1} R^0_j \right)
\label{P0}
\end{equation}
(Summation over repeated indices has been assumed).  The matrix
$\ell^0_{ij}$ of effective step lengths defines  the chain shape
parallel and perpendicular to the  director ${\bf n}^0$ for a
uniaxial phase, that is
\begin{equation}
\langle R^0_i R^0_j \rangle =
\frac{1}{3}{\cal L} \ell^0_{ij}  \ , \label{RR}
\end{equation}
 where  ${\cal L}$ is the chain contour
length. The effective step lengths of the random walk, and thus
the overall average shape, are functions of the nematic order
parameter.  For a uniaxial nematic one can write:
\begin{equation}
\ell_{ij} = \lp \delta_{ij} + (\lz - \lp) \, n_i \, n_j \ \ ,
\label{lij}
\end{equation}
where the explicit functional form $\lz(Q)$ and $\lp(Q)$ is
dependent on the particular model of the liquid crystal polymer
chain. We shall not need this information to describe the
rubber-elastic effects.

Next comes the affine deformation
assumption, an assumption that pervades network
theory. Junction point fluctuations are  damped by
connectivity and in the case of very high crosslink functionality
compel crosslinking points to move exactly in geometric proportion
to the whole sample.  We use the simplifying but unnecessary
affine deformation assumption and define the current network span
to be $R_i=\lambda_{ij}R^0_j$ with $\lambda_{ij}$  the macroscopic
(Cauchy) strain tensor of the whole block of rubber.
The polymer strand end-to-end vector probability is
given by a distribution, $P({\bf R})$, differing from
$P_0({\bf R}^0)$ in Eq.(\ref{P0}) in that the $(\ell_{ij})^{-1}$
tensor, describing the current chain shape, depends on the
{\it current} state of the nematic order after the deformation
has taken place. Taking  the usual quenched average for each
network strand $F_{el}=- \langle k_BT\ln P({\bf
R})\rangle_{P_0({\bf R}^0)}$ ,  one obtains the elastic free
energy density describing uniform deformations of liquid crystal
elastomers \cite{our-hard}:
\begin{eqnarray}
F_{el}  = \frac{1}{2} N_x k_BT \ {\sf Tr}[\, \matr{\ell}^0
\cdot \matr{\lambda}^T \cdot \matr{\ell}^{-1}
\cdot \matr{\lambda}  \, ] \   ,  \label{Fel}
\end{eqnarray}
where $N_x$ is the number of crosslinks per unit volume (see
\cite{WGV} for insignificant corrections due to the junction point
fluctuations) and we have used $\langle R^0_i R^0_j \rangle =
\frac{1}{3}{\cal L} \ell^0_{ij}$, Eq.(\ref{RR}). There is also an
additional term which  arises from the normalization of $P({\bf
R})$ (see \cite{WGV,our-Q}). This term depends on the magnitude of
the nematic order parameter $Q$, which could, in principle, also
be affected by elastic strains. At sufficiently low temperatures,
away from the clearing point $T_{ni}$, the degree of polymer
chains anisotropy $\lz/\lp$ is strongly constrained by
thermodynamic forces and it is quite safe to assume $Q=const$ and
consider only the rotations of the nematic director \vec{n} in
response to elastic strains in Eq.(\ref{Fel}), so that
$\ell_{ij}$ would be just a rotated version of the matrix
$\ell_{ij}^0$ (see the review article \cite{our-review} for a
detailed discussion).

The elastic free energy (\ref{Fel}) showns a rich behaviour,
based on the fact that the elastic strain tensor enters this
expression in a generally non-symmetric form, unlike in any other
solid material (or isotropic rubber), where one always has
$F_{el} \sim \{ \matr{\lambda}^T \cdot \matr{\lambda} \}$.
Antisymmetric components of strain couple to the director
rotation away from $\vec{n}_0$ (given by $\matr{\ell}^0$) to
\vec{n} (the principal axis of the current step length tensor
$\matr{\ell}$). In particular, after implementing the limit of
infinitesimal deformations $\matr{\lambda} = \vec{1} +
\matr{e}$, one obtains molecular expressions for de Gennes'
coupling terms in Eq.(\ref{rotations}):
\begin{equation}
b_1 = N_x k_BT \, \frac{(\lz-\lp)^2}{\lz \lp} \, ;
\ \ \  b_2 = N_x k_BT \, \frac{\lz^2-\lp^2}{\lz \lp}
\nonumber
\end{equation}
(a positive coupling $b_2$ corresponds to prolate
polymers,those with  mesogenic
units in or aligned parallel to the backbone and thus with $\lz >
\lp$). Several new physical effects have been predicted with the
help of Eq.(\ref{Fel}), notably discontinuous nematic transitions
driven by an imposed elastic strain (also observed experimentally
\cite{mitch-prl}), reorientation by external electric or magnetic
fields and the so-called `soft elasticity'.  However, the  above
arguments are applicable only in the case of uniform director
rotations and uniform elastic strains. Applications of
Eq.(\ref{Fel}), however appealing and powerful they might look,
are severely hindered by its neglect of curvature
deformations which are quite common, even in elastomers, due to
surface anchoring, disclinations and domain walls in equilibrium,
or during orientational transitions.

It is the purpose of this paper to derive the non-uniform elastic
free energy, by a modification of Eq.(\ref{Fel}), that
describes the coupling between curvature deformations and elastic
strains in nematic elastomers.

\section{Non-uniform director field }

We shall assume, as a simplifying starting point, that the
director distribution before deformation, $\vec{n}_0$, which is
implicitly present in the initial chain step length tensor
$\ell_{ij}^0$, is uniform. In other words, the polymer network
has been formed in a uniform monodomain nematic state, or
brought to its present state in a sufficiently strong external
field. Another simplifying assumption will be that the present
temperature is sufficiently below the clearing point $T_{ni}$, so
that the magnitude of the nematic order (and the related chain
step length anisotropy $\lz/\lp$) is not changed during the
deformation and only director rotation takes place.
Both these assumptions are not crucial for the arguments below
and a straightforward generalization to account to these effects
is possible. However, these two factors, $\nabla \vec{n}_0$
and $\ell_{ij}(Q)$, bring additional degrees of freedom and
geometric constraints, which make the resulting elastic free
energy very difficult to read. We prefer to present the main
line of derivation in a more clear, albeit slightly less general
form and, therefore, consider $\vec{n}_0 = {\rm const}$ and
$\lz/\lp = {\rm const}$.

In order to determine the elastic response of a random polymer
network to a non-uniform deformation field  one has to find
the statistical weight $P(R)$ of configurations of a given
polymer strand in the presence of such a field. This statistical
weight for a chain with a fixed contour length, ${\cal L}$, is
determined by the general tensor of chain persistent lengths
$\ell_{ij}$, see Eqs.(\ref{P0}), (\ref{Fel}). Then the quenched
averaging should be performed, exactly as in Eq.(\ref{Fel}), with
respect to the initial state $P(R^0)$, which is characterised by
the corresponding initial  $\ell_{ij}^0$ and which we assumed to
be uniform.  Therefore, we need to describe the configurations of
a liquid crystal polymer chain in the presence of director
curvature deformations $\ell_{ij}(\vec{n}, \nabla \vec{n})$.
After expansion in powers of small gradients
(long-wavelength limit) we should then have in a uniaxial
non-chiral nematic:
\begin{equation}
\ell_{ij} \approx \ell_{ij}^{[u]} + \kappa_{ijabcd}(\vec{n})
\nabla_a n_b \nabla_c n_d + . . .   \ ,  \label{l-1mod}
\end{equation}
which tells us that the shape of the polymer coil is slightly
altered when the mesogenic units of this chain are
subjected to a spatially non-uniform nematic mean field
potential ($\ell_{ij}^{[u]}$ represents the uniaxial step
length tensor of the uniform system). The statistical weight of
such a chain with its end-to-end vector $\vec{R}$ affected by the
affine elastic deformation of the network is, as before,
\begin{equation}
P({\bf R}) \sim   \exp \left( - \frac{3}{2{\cal L}}
\{ \matr{\lambda} \cdot R^0\}_i \, \ell_{ij}^{-1}(\vec{n}, \nabla
\vec{n}) \, \{\matr{\lambda} \cdot R^0\}_j \right)  \ .
\label{PP}
\end{equation}
Performing the quenched averaging with the (uniform) initial
probability distribution $P(\vec{R}^0)$, Eq.(\ref{P0}), one
obtains
\begin{eqnarray}
F_{el}  &=& \frac{1}{2} N_x k_BT \ {\sf Tr}[\, \matr{\ell}^0
\cdot \matr{\lambda}^T \cdot \matr{\ell}^{-1}
\cdot \matr{\lambda}  \, ]    \nonumber    \\
&\approx & F_{el}^{[u]}  -
\frac{1}{2} N_x k_BT \ {\sf Tr} [\, \matr{\ell}^0
\cdot \matr{\lambda}^T \cdot \{ ... \nabla \vec{n}
\nabla \vec{n} ... \} \cdot \matr{\lambda}  \, ]  \ ,
\label{Fmodel}  \end{eqnarray}
where $F_{el}^{[u]}$ is the uniform liquid crystal elastomer
elastic energy Eq.(\ref{Fel}). Note the minus sign at the
non-uniform part of $F_{el}$, which is due to the inversion
$\ell^{-1}$ from Eq.(\ref{l-1mod}) and which will appear to
represent a genuinely negative free energy
contribution. \\

How can one find  $\ell_{ij}(\vec{n}, \nabla \vec{n})$? Obviously
a specific model of the liquid crystal polymer chain must be
employed and the result could be very different for
main-chain mesogenic polymers (best modelled by a persistent
worm-like chain) and for side-chain materials (their backbone
configuration is adequately described by the much simpler
freely-jointed rod model). Since the majority of existing
liquid crystal elastomers are made of side-chain polymers, we
shall concentrate on this case. We present a calculation of the
average end-to-end distance $\langle R_i R_j \rangle$ of the
uniaxial nematic polymer in the non-uniformly distorted director
field \vec{n}(\vec{r}), within a freely-jointed chain model.
Later we shall also briefly outline the similar calculation for
the main-chain work-like nematic polymer.

Let us choose the starting point of the chain trajectory as the
origin of coordinate system, then the position of the
$\alpha^{th}$ monomer on the chain is given by $\vec{r}^{\{\alpha
\}} =  a \sum_{\nu=1}^{\alpha} \vec{u}^{\{\nu\}} $, where $a$ is
the physical length of the monomer and $\vec{u}_\nu$ is the
tangent vector of the monomer number $\nu$ on this chain. The
direct product of end-to-end vectors is, therefore, $\vec{R} \,
\vec{R} =  a^2 \sum_{\nu,\nu' = 1}^{N} \vec{u}^{\{\nu\}}
\vec{u}^{\{\nu' \}} $ , where $N = {\cal L}/a $ is the number of
monomers on the chain. For the freely-jointed chain there is no
correlation between orientations of different monomers,
$\vec{u}^{\{\nu\}} $ and $\vec{u}^{\{\nu' \}} $ for $\nu \neq
\nu'$, and the shape of the polymer coil is determined by
\begin{eqnarray}
\langle R_i R_j \rangle &=& a^2 \sum_{\nu=1}^N \langle
u_i^{\{\nu\}} u_j^{\{\nu\}} \rangle   \label{free-RR}    \\
&=& \frac{1}{3} a N \sum_{\nu=1}^N \bigg( \lp \delta_{ij} +
[\lz - \lp] n_i(\vec{r}^{\{\nu\}}) n_j(\vec{r}^{\{\nu\}}) \bigg)
\nonumber
\end{eqnarray}
(we take Latin indices to represent the vector components, $i,j =
1,2,3$, while Greek indices number monomers along the chain,
$\alpha, \nu = 1, ... \, , N$. For the freely-jointed chain
there is a trivial relation between the monomer length, step
lengths and the backbone order parameter $Q$: $a =
\frac{1}{3}(\lz + 2 \lp)\, ; \ Q = \frac{1}{3} (\lz - \lp)/a$.
In the second expression in Eq.(\ref{free-RR}) we explicitly
expose the fact that the orientation of the principal axis of
the uniaxial average $\langle \vec{u} \vec{u} \rangle$ depends on
the position of the corresponding monomer, numbered $\nu$.
This is the crux of the problem since $\vec{r}^{\{\nu\}}$ itself
depends on all the preceeding monomers orientations and hence on
the nematic director \vec{n}(\vec{r}) on all these locations.
The formerly simple problem of a freely-jointed random walk has
become a higher order Markov process, the $\nu^{th}$ step
depending on all the previous steps. This problem can be solved
systematically in powers of $(\nabla \vec{n})^2$.

The end-to-end vector \vec{R} of a chain with $N$ monomers can
also be written in a recurrent form, $\vec{R} =
\vec{r}^{\{N-1\}} + a\vec{u}^{\{N\}}$ and thus the average
required in Eq.(\ref{free-RR}) is
\begin{equation}
\langle R_i R_j \rangle =
\langle r_i^{\{N-1\}} r_j^{\{N-1\}} \rangle + a^2\langle
u_i^{\{N\}} u_j^{\{N\}} \rangle
\label{A1}
\end{equation}
(cross terms $\langle r^{\{N-1\}} u^{\{N\}} \rangle$ vanish in
a non-chiral system for obvious symmetry reasons). The last term
in (\ref{A1}) is equal to $\frac{1}{3}a \big( \delta_{ij} \lp +
[\lz - \lp] n_i(\vec{r}^{\{N-1\}}) n_j(\vec{r}^{\{N-1\}}) \big)$
where it is explicitly noted that the directions of the $N^{\rm
th}$ link are biased according to the director $\vec{n}
(\vec{r}^{\{N-1\}})$ at its starting point $\vec{r}^{\{N-1\}}$.
 Recognizing that in the
long-wavelength limit the difference between
$\vec{n}(\vec{r}^{\{1\}}) $ and $\vec{n}(\vec{r}^{\{N\}}) $ is
small (which implies the upper cut-off for the curvature wave
vectors, $|\vec{q} | \ll |\vec{R} |^{-1} $, the inverted network
mesh size), we perform the gradient expansion:
$n_i(\vec{r}^{\{N-1\}}) \rightarrow \  n_i(\vec{r}^{\{1\}}) +
(\vec{r}^{\{N-1\}} \cdot \nabla) n_i(\vec{r}^{\{1\}})
+ \frac{1}{2} (\vec{r}^{\{N-1\}} \vec{r}^{\{N-1\}} : \nabla
\nabla ) n_i(\vec{r}^{\{1\}}) + ...$ , where one should treat
$n_i(\vec{r}^{\{1\}})$ as the local current director \vec{n}.
Substituting this into Eq.(\ref{A1}) we obtain
\begin{eqnarray}
\langle R_i R_j \rangle &=&
\langle r_i^{\{N-1\}} r_j^{\{N-1\}} \rangle +
\frac{1}{3}a \big( \lp \delta_{ij} +   [\lz - \lp] n_i n_j
\big)        \label{A2} \\
&& \qquad + \frac{1}{3}a [\lz-\lp] \langle r_k^{\{N-1\}}
r_l^{\{N-1\}}  \rangle \left[(\nabla_k n_i)(\nabla_ln_j) +
\frac{1}{2} n_i \nabla_k\nabla_l n_j +
\frac{1}{2} n_j \nabla_k\nabla_l n_i \right]   \nonumber
\end{eqnarray}
(we have eliminated cross terms like $\langle n_i (
\vec{r}^{\{N-1\}} \cdot \nabla n_j) \rangle$ which vanish for a
non-chiral system). So far Eq.(\ref{A2}) is exact at
$\underline{\cal O}(\nabla \vec{n})^2$  and the problem of
correlations has been set back to $\langle r_i^{\{N-1\}}
r_j^{\{N-1\}} \rangle$. This equation can now be iterated to
give $\langle r_i^{\{N-1\}} r_j^{\{N-1\}} \rangle$ in terms of
$\langle r_k^{\{N-2\}} r_l^{\{N-2\}} \rangle$ and gradients of
\vec{n}, and so on. Each step of such iteration generates
successively higher powers of  $[\lz-\lp](\nabla \vec{n})^2$.
Since there are no linear terms, ignoring all powers greater
than $(\nabla \vec{n})^2$ is equivalent to taking the
uniform-\vec{n} value for $\langle r_k^{\{\nu\}} r_l^{\{\nu\}}
\rangle$ [see Eq.(\ref{free-RR}) with $N=\nu$]. In this way we
obtain
\begin{eqnarray}
\ell_{ij} &\equiv & \frac{3}{aN} \langle R_i R_j \rangle =
\lp \delta_{ij} +   [\lz - \lp] n_i n_j     \label{lgrad} \\
&& + \frac{1}{6}a N [\lz - \lp] \bigg[
(\lz - \lp) (\vec{n}\cdot \nabla n_i) (\vec{n}\cdot \nabla n_j) +
\lp  (\nabla_k n_i) (\nabla_k n_j)     \nonumber \\
&& \qquad \qquad + \frac{1}{2}(\lz - \lp)
[n_i (\vec{n}\vec{n}:\nabla \nabla) n_j +
n_j (\vec{n}\vec{n}:\nabla \nabla) n_i]  +
\frac{1}{2} \lp [n_i \nabla^2 n_j + n_j \nabla^2 n_i] \bigg]
\nonumber
\end{eqnarray}
where the extra power of $N$ has appeared due to the summation
$\sum_{\nu=1}^N \nu = \frac{1}{2}N(N+1)$ of all
$\langle r_i^{\{\nu\}} r_j^{\{\nu\}} \rangle$ terms in the
iterated Eq.(\ref{A2}). Inverting the matrix (\ref{lgrad}) is
trivial because we should only keep the lowest (second in this
case) powers of $\nabla \vec{n}$; in this limit of small
gradients we obtain \begin{equation}
\ell_{ij}^{-1} \approx \ell_{ij}^{[u]} {}^{-1} -
\frac{1}{6}a N (\lz-\lp) \, \ell_{im}^{[u]} {}^{-1}  \,
\bigg[ \ . \ . \ . \  \bigg]_{mk} \, \ell_{kj}^{[u]} {}^{-1}
\ ,  \label{l-1}
\end{equation}
where $\ell_{ij}^{[u]} $ is the local ``uniform'' step length
tensor  depending on \vec{n}, and the expression in square
brackets should be directly taken from Eq.(\ref{lgrad}).

All that remains is to substitute this inverse of the non-uniform
step length tensor in the general local rubber-elastic free
energy density Eq.(\ref{Fmodel}) and integrate by
parts in order to convert the result to the form consistent with
$\{ \nabla \vec{n} \}^2 $ only [instead of having the
second-derivative terms, suggested by the (\ref{lgrad})]. This
integration by parts is quite tedious and strictly speaking,
since the elastic strain tensor is in general a function of
coordinates, will generate derivatives of $\lambda$ too. However,
it is common in elasticity theories to neglect the gradients of
strains (which correspond to second derivatives of displacement
in infinitesimal models). We, accordingly, shall ignore the terms
of the form $\{ \nabla \vec{\lambda}^2  \ \nabla \vec{n} \} $ in
favour of the coupling to the uniform part of strains $\{
\vec{\lambda}^2 \, (\nabla \vec{n} )^2 \}$ , which bears a
superficial similarity with the nematic Frank free energy and
rubber-elastic energy (\ref{Fel}). In this case the algebra is
straightforward and we obtain the main result of this paper, the
elastic free energy density of non-uniform director deformations
in nematic  elastomers [compare with Eq.(\ref{Fmodel})]:
\begin{eqnarray}
\Delta F_{el} &=& - \frac{1}{2} \kappa_1 \ {\sf
Tr} \left[\, \matr{\ell}^0 \cdot \matr{\lambda}^T \cdot
\big\{ (\vec{n}\cdot \nabla)n_i \, (\vec{n}\cdot \nabla)n_j
\big\} \cdot \matr{\lambda}  \, \right]     \label{mainF}  \\
&&  - \
\frac{1}{2} \kappa_2 \ {\sf Tr} \left[\,
\matr{\ell}^0 \cdot \matr{\lambda}^T \cdot
\big\{ \nabla_k n_i \nabla_k n_j \big\}
\cdot \matr{\lambda}  \, \right]       \nonumber  \\
&& - \ \frac{1}{2} \kappa_3 \ {\sf Tr} \left[\,
\matr{\ell}^0 \cdot \matr{\lambda}^T \cdot
\{ n_i n_j \}
\cdot \matr{\lambda}  \, \right] \, [(\vec{n}\cdot \nabla)
\vec{n}]^2 \nonumber \\
&& - \ \frac{1}{2} \kappa_4 \ {\sf Tr} \left[\,
\matr{\ell}^0 \cdot \matr{\lambda}^T \cdot
\{ n_i n_j \}
\cdot \matr{\lambda}  \, \right] \, (\nabla_k n_l)(\nabla_k n_l)
\nonumber   \\
&& + \
\frac{1}{2} \kappa_5 \ \bigg( {\sf Tr} \left[\,
\matr{\ell}^0 \cdot \matr{\lambda}^T \cdot
\big\{ n_i(\vec{n}\cdot \nabla)n_k \nabla_k n_j + n_j(\vec{n}
\cdot \nabla)n_k \nabla_k n_i \big\} \cdot \matr{\lambda}  \,
\right]  \nonumber \\
&& \qquad \qquad +
{\sf Tr} \left[\,
\matr{\ell}^0 \cdot \matr{\lambda}^T \cdot
\big\{ n_i(\vec{n}\cdot \nabla)n_j + n_j(\vec{n}\cdot
\nabla)n_i \big\} \cdot \matr{\lambda}  \, \right] \, div \,
\vec{n}   \bigg)  \nonumber   \ ,
\end{eqnarray}
where the elastic constants have the form
\begin{eqnarray}
\kappa_1 &=& \frac{1}{6}\rho k_BT \, a \frac{(\lz-\lp)^3}{\lz
\lp^2} \ ; \qquad \kappa_2 = \frac{1}{6} \rho k_BT \, a
\frac{(\lz-\lp)^2}{\lz \lp}  \ ;  \label{const-1} \\
\kappa_3 &=& \frac{1}{6}\rho k_BT \, a \frac{(\lz-\lp)^3}{\lz^2
\lp} \ ; \qquad \kappa_4 = \frac{1}{6} \rho k_BT \, a
\frac{(\lz-\lp)^2}{\lz^2} \ ;   \nonumber  \\
\kappa_5 &=& \frac{1}{12}\rho k_BT \, a \frac{(\lz-\lp)^2}{\lz
\lp}  \ ,   \nonumber
\end{eqnarray}
and where $\rho = N_x N$ is the total number density of monomers
in the system.

There can be different representations of these elastic
constants, using chain step lengths, physical monomer length, or
the backbone order parameter. Experimentally, perhaps the most
easily accessible parameters of the material are the monomer size
$a$ (simply from its chemical structure) and the aspect
ratio of the nematic polymer chain $\Delta = \lz/\lp$ (from
neutron scattering or from the sample shape change on
isotropization), which is also the single parameter of the
``uniform'' theory \cite{our-hard} and determines all elastic
instabilities. In these variables, the five non-uniform
nematic rubber-elastic constants take the form:
\begin{eqnarray}
\kappa_1 &=& \frac{1}{2}\rho k_BT \, a^2 \frac{(\Delta -
1)^3}{\Delta (\Delta+2)} \ ; \qquad \kappa_2 = \frac{1}{2} \rho
k_BT \, a^2 \frac{(\Delta -1)^2}{\Delta (\Delta + 2)}  \ ;
\label{const-2} \\
\kappa_3 &=& \frac{1}{2}\rho k_BT \, a^2
\frac{(\Delta -1)^3}{\Delta^2 (\Delta + 2)} \ ; \qquad \kappa_4
= \frac{1}{2} \rho k_BT \, a^2 \frac{(\Delta - 1)^2}{\Delta^2
(\Delta+2)} \ ;   \nonumber   \\
 \kappa_5 &=& \frac{1}{4}\rho k_BT
\, a^2 \frac{(\Delta -1)^2}{\Delta (\Delta+2) }  \ .   \nonumber
\end{eqnarray}
In order to complete these equations, an additional
factor of $\lp$ has been pulled out of the tensor $\matr{\ell}^0$
in the traces  in  Eq.(\ref{mainF}), so that the initial step
length tensor $\ell_{ij}^0$ there should now be regarded as
dimensionless, $\ell_{ij}^0 = \delta_{ij} +  (\Delta -1)\, n_i^0
n_j^0$. For freely jointed rod polymers $\lp = 3a/(\Delta + 2)$.
One can immediately notice that the dimensionality of the elastic
constants $\kappa$ is energy per length, the same as the Frank
constants of a nematic.

The reader should be reminded that the above derivation is
performed explicitly within the framework of freely-jointed
rod model for the polymer chain. In the most general case
there are two more non-uniform nematic rubber elastic terms,
arising from the persistent correlations along the chain.
\begin{eqnarray}
 \ - \frac{1}{2}\kappa_6 {\sf Tr} \left[\,
\matr{\ell}^0 \cdot \matr{\lambda}^T \cdot
\{ \delta_{ij} \}
\cdot \matr{\lambda}  \, \right] \, [(\vec{n}\cdot \nabla)
\vec{n}]^2 \ - \ \frac{1}{2}\kappa_7
 {\sf Tr} \left[\,
\matr{\ell}^0 \cdot \matr{\lambda}^T \cdot
\{ \delta_{ij} \}
\cdot \matr{\lambda}  \, \right] \,(\nabla_k n_l)(\nabla_k n_l)
\label{addF}
\end{eqnarray}
The effect of these two terms is not qualitatively different from
the terms $\kappa_3$ and $\kappa_4$ of the main
Eq.(\ref{mainF}) and, since they depend only on the symmetric
product of strains $\matr{\lambda}^T \cdot \matr{\lambda}$ , no
new nematic effects should be expected from these particular
terms.  We did not specifically endeavor
to obtain the molecular expressions for the corresponding
constants and, for all practical purposes, we should
adopt the approximate equations for the elastic constants
(\ref{const-2})  of the general non-uniform nematic
rubber-elastic free energy.

\section{Discussion}

{}From first glance, the non-uniform elastic energy of a
nematic rubber attracts attention by its negatively
determined structure. One could, perhaps, intuitively accept this
qualitative form, $-\kappa \, \lambda^2 (\nabla \vec{n})^2$ ,
as demanding an elastic distortion of the sample, subjected to a
curvature deformation of the director. One, however, should be
careful with conclusions because the elastic free energy
(\ref{mainF}) contains several matrix products and its overall
scalar magnitude depends on the mutual orientation of several
independent objects: the initial director $\vec{n}^0$ before
deformation, the elastic strain tensor, the local director
\vec{n} and its gradient. In addition, the effect of $\Delta
F_{el}$ must be considered together with two other relevant
contributions, the ``uniform'' nematic rubber elasticity
$F_{el}^{[u]}$, Eq.(\ref{Fel}), which penalizes any deviations
of the director with respect to initial $\vec{n}^0$, and the
underlying, conventional Frank nematic elasticity \cite{book},
which is the response to any non-uniform deformation $\nabla
\vec{n}$. In order to examine the implications of this new elastic
energy we should consider some simple particular cases.

In the simplest imaginable situation one
clamps the sample, thus prohibiting all uniform
deformations. (There would still be a possibility for the soft
material to deform in a non-uniform fashion with, say, $\int
[ \lambda(z) - 1] dz = 0$, but such an effect could be neglected
in the first approximation as being of higher order in small
deformations). Taking $\lambda_{ij} = \delta_{ij}$ we arrive at
the following expression
\begin{eqnarray}
\Delta F_{el} &=& -\frac{1}{2}\big[ \kappa_1 + \kappa_3 [1 +
(\Delta-1) (\vec{n}^0 \cdot \vec{n})^2] +
(\Delta+2)\kappa_6 \big] \, [(\vec{n}\cdot \nabla) \vec{n}]^2
 \label{clampF} \\
&& -\frac{1}{2}\big[ \kappa_2 + \kappa_4 [1 + (\Delta-1)
(\vec{n}^0 \cdot \vec{n})^2] +  (\Delta+2)\kappa_7 \big] \,
(\nabla_k n_l)(\nabla_k n_l)   \nonumber  \\
&& - \frac{1}{2} \kappa_1 (\Delta-1) (\vec{n}^0 \cdot
(\vec{n}\cdot \nabla) \vec{n})^2   - \frac{1}{2} \kappa_2
(\Delta-1) (\vec{n}^0 \cdot \nabla_k \vec{n})
 (\vec{n}^0 \cdot \nabla_k \vec{n})
\nonumber \\
&&  +  \kappa_5 (\Delta-1) (\vec{n}^0\cdot \vec{n})
\big[ \vec{n}^0 \cdot (\vec{n}\cdot \nabla n_k) \nabla_k \vec{n}
+  \vec{n}^0 \cdot (\vec{n}\cdot \nabla \vec{n}) \, div \,
\vec{n} \big]
\end{eqnarray}
Here, in contrast to simple nematics, the difference between
$\vec{n}^0$ and $\vec{n}$ is a director rotation that is penalised
even for uniform distortions.  The energetic cost is substantial,
of the order of $\mu$ per unit volume and  it is thus very likely
that such a rotation cannot be easily achieved, for example the
influence of external electric or magnetic fields is shown
\cite{our-El} to be quite insufficient to create a significant
deviation of $\vec{n}$ from $\vec{n}^0$. If we then take  $\vec{n}
= \vec{n}^0 + \delta \vec{n}(\vec{r})$ and retain, as usual, only
the leading terms in small non-uniform deviations $\delta
\vec{n}$, the elastic energy simplifies dramatically to:
\begin{eqnarray}
\Delta F_{el} &=&
 -\frac{1}{2}\big( \kappa_2 + \kappa_4 \Delta
+  [\Delta+2]\kappa_7 \big) \,  (div \, \vec{n})^2
-\frac{1}{2}\big( \kappa_2 + \kappa_4 \Delta
+  (\Delta+2)\kappa_7 \big) \,
(\vec{n} \cdot \nabla \times \vec{n})^2       \label{clamp2}  \\
&& \qquad -\frac{1}{2}\big( \kappa_1 + \kappa_2 + (\kappa_3  +
\kappa_4)\Delta  + [\Delta+2](\kappa_6 + \kappa_7) \big) \,
[(\vec{n}\times \nabla \times \vec{n}]^2    \ ,
\nonumber
\end{eqnarray}
which has an exact form of
the Frank elasticity of regular nematic liquid crystals. This is
an expected conclusion, since we have eliminated all independent
vector and tensor variables from our simplified, clamped system
leaving only gradients of \vec{n}. What can be considered
unexpected is that the remaining combinations of elastic constants
yield  negative square gradient contributions to the free energy!

It is important to compare the magnitude of the new elastic
constants with the standard Frank constants $K_{11} \, , K_{22}$
and $K_{33}$, which work against the above destabilizing effect.
The order of magnitude of our new constants is sufficiently easy
to estimate because thay are determined by the usual entropic
effects of  polymers in  rubber
elasticity. Within  orders of magnitude  we have \ $\kappa \sim
\rho k_BT a^2$ or, in  dense thermotropic side-chain systems,
$\kappa \sim k_BT/a \ \sim 10^{-11} \,  J/m$. This is comparable
to values of typical Frank constants. If one considers a
main-chain polymer liquid crystal, with an extremely large
backbone anisotropy ratio $\Delta = \lz/\lp \gg 1$, forming a
rubbery network, this comparison can become even more favourable
to the  new negative non-uniform nematic rubber elasticity.

As we have discussed above, this new elastic energy should be
compared with the ``uniform'' nematic rubber elasticity,
Eq.(\ref{Fel}), which produces for the same clamped case
\begin{equation}
F_{el}^{[u]} = \frac{1}{2} N_x k_BT \frac{(\Delta -1)^2}{\Delta}
\, \big( \delta \vec{n} \big)^2    \ , \label{clampU}
\end{equation}
{\it i.e.} the magnitude of the director rotation, whether
uniform or non-uniform, is penalized by the anchoring to the
clamped network \cite{our-hard,our-El}. Let us assume that the
destabilizing effect of Eq.(\ref{clamp2}) has won against the
Frank elasticity and consider the overall orientational stability
of the sample. For a qualitative analysis we take the angular
deviation $\delta \vec{n} \sim \theta_0 \cos qx$ and write both
the uniform and non-uniform parts of the elastic energy, pulling
out common factors:
\begin{equation}
F_{el} \sim \frac{1}{2} N_x k_BT \theta_0^2 \, \big[ 1 -
N a^2 q^2 \big]  \label{instab}
\end{equation}
(we have neglected all other factors, like $\Delta$, which we
assume to contribute factors of order unity in a side-chain
polymer system). Within Eq.(\ref{instab}), clearly, the
orientational instability may occur only with very short  wave
lengths,  $q^{-1} \sim a\, N^{1/2} $, which is of the order of
network mesh size ($\langle R^2 \rangle \sim a^2 N$). A more
detailed analysis of higher order contributions, {\it i.e.}
$(\nabla \vec{n})^4, \ (\nabla \vec{n})^6$, etc., would then be
required. If such small
textures were to be formed (in a typical experimental case  $a \,
N^{1/2} \sim 30-50 \, \AA$), they would be impossible to detect
by any optical method and the nematic elastomer would appear
uniform. It is likely therefore that in the equilibrium state of
a nematic elastomer spontaneous long wave-length
deformations, which would be favoured by the new elastic energy
$\Delta F_{el}$, are prohibited by the strong uniform anchoring
of the director to the network.

This situation may be changed in three circumstances.

\noindent (i) In a mechanically unconstrained sample, or
one with non-uniform distortions such that the sample is globally
undistorted,  there exists a possibility of having so-called
`soft elastic deformations' \cite{our-soft,olmsted}. These
do not give rise to the uniform part of elastic energy
$F_{el}^{[u]}$, and happen when a special class of elastic
strains is combined with certain director rotations in such a way
that the network polymer strands are not forced to change their
equilibrium shape.  These strains therefore do not cause  a drop
in configurational entropy. One example of such a soft strain
tensor could be $\matr{\lambda} = \matr{\ell}^{1/2}
\matr{\ell}_0^{-1/2}$ \ (see \cite{olmsted} for details).
Insertion of $F_{el}^{[u]}$ of Eq.(\ref{Fmodel}), that is
Eq.(\ref{Fel}) with uniform fields, shows trivially that this
class of strains leaves the ``uniform'' elastic energy unchanged
from the unstrained state. It can  still allow a negative non
uniform contribution $\sim -\kappa (\nabla \vec{n})^2$. It is
expected that networks crosslinked in isotropic state and then
cooled down into the nematic phase should exhibit such `soft
elasticity' due to their rotational invariance \cite{GL}.  The
destabilising negative gradient terms, unrestrained by the
positive penalty for uniform rotation, may explain why all such
elastomers in practice form scattering polydomain textures in
thermodynamic equilibrium.

\noindent (ii)  Another, perhaps more relevant case, when the
destabilising effect of the new non-uniform elastic energy
$\Delta F_{el}$ can be felt is during the mechanically driven
orientational transitions, for example the ones described in
\cite{mitch-prl,our-hard}. An imposed elastic strain overcomes
the barrier to director rotation, given by the uniform energy
$F_{el}^{[u]}$.  The transitions predicted and observed were to a
uniform, rotated state.  But we now see that orientational
modulations could occur in the strained state. This, of course
should take place only when the new constants $\kappa$ are
actually more relevant than the (stabilizing) Frank constants. We,
therefore,  predict that materials with higher backbone
anisotropy $\Delta = \lz/\lp$ should be more likely to exhibit
such spontaneous breaking into orientational domains during
various director transitions.

\noindent (iii)  It is possible, when there is sufficiently
strong surface anchoring of the director in a direction in
conflict with the principal axes of the imposed strain, that one
can have a mechanical Freedericks transition to a non uniform
state.  Even if the new, negative constants do not outweigh the
Frank constants in the limit  Eq.(\ref{clamp2}), it is
possible that for larger $\lambda$ and for $\vec{n}$ not close to
$\vec{n}^o$, these effects help the nematic rubber elastic
contributions overcome the Frank penalty and tip the balance in
favour of the distorted state with non-uniform nematic textures.
\vspace{0.4cm}

We appreciate many stimulating discussions with P.D. Olmsted and
H. Finkelmann. This research has been supported by EPSRC (GV)
and by Unilever, PLC (EMT and MW).


\end{document}